\documentclass{article}
\usepackage{iclr2027_conference,times}
\usepackage{graphicx,booktabs,amsmath,amssymb,longtable,placeins}
\usepackage{algorithm,algpseudocode}
\usepackage{hyperref}
\usepackage{url}
\title{Learning as Deepfakes Evolve: RF-Prompt for Continual Audio Deepfake Detection}
\author{%
\small\textbf{Yuankun Xie$^{1}$, Xiaoxuan Guo$^{2}$, Xiaopeng Wang$^{3}$, Siqing Qin$^{1}$, Shaole Li$^{1}$, Kong Aik Lee$^{1}$}\\[3pt]
\normalfont $^{1}$The Hong Kong Polytechnic University, Hong Kong SAR, China\\
\normalfont $^{2}$Communication University of China, Beijing, China\\
\normalfont $^{3}$Beijing Institute of Technology, Beijing, China
}
\iclrfinalcopy
\hypersetup{%
  hidelinks,
  pdfauthor={Yuankun Xie, Xiaoxuan Guo, Xiaopeng Wang, Siqing Qin, Shaole Li, Kong Aik Lee},
  pdftitle={Learning as Deepfakes Evolve: RF-Prompt for Continual Audio Deepfake Detection}
}
\begin{document}
\maketitle

\begin{abstract}

Speech generation methods are evolving rapidly, creating a moving target for audio deepfake detection (ADD). A deployed detector must incorporate newly emerging deepfake methods without forgetting previously learned real and deepfake knowledge. Continual learning provides a natural solution, but existing continual ADD evaluations commonly define tasks by dataset, coupling changes in real-speech sources with changes in deepfake mechanisms and obscuring what knowledge is being updated. We address task organization and detector adaptation jointly. First, we construct five protocols from identical training, development, and evaluation pools. Among them, the proposed real-anchored mechanism-incremental (RAMI) protocol reflects a practical detector-update scenario: real speech from known source domains is available, while newly arriving deepfake mechanisms must be learned continually without forgetting earlier knowledge. Second, we propose RF-Prompt (real--fake prompt learning), an asymmetric continual prompt-learning method. A shared real prompt provides protected adaptation capacity, while task-specific fake experts expand as new mechanisms arrive. Parameter-level cosine anchoring stabilizes the real prompt; each new fake expert inherits a selected historical expert and learns a residual with soft orthogonal regularization. Input-adaptive fusion integrates the accumulated fake experts without requiring task identity at inference. RF-Prompt obtains its lowest common-average and pooled EER under RAMI among the five tested protocols. On RAMI, it achieves 10.110\% final average EER and 10.370\% pooled EER, the lowest aggregate errors among the evaluated continual-learning baselines. Ablations assess the adaptation components, while limited-training-data and cross-backbone experiments examine their applicability across settings.
Code is available online\footnote{\url{https://github.com/xieyuankun/RF-Prompt}}.
\end{abstract}
\section{Introduction}

Speech generation is becoming increasingly accessible, expressive, and diverse. Industrial systems such as Alibaba's Qwen3-TTS, ByteDance's Seed-TTS, OpenAI's GPT-Live, and Google's Gemini Live are evolving rapidly, with new models and updates appearing on monthly or even weekly cycles \citep{qwen3tts,seedtts,gptlive,gemini38live}. This rapid evolution creates a moving target for audio deepfake detection (ADD): a deployed detector must continually confront generators that were unavailable during training. Reliable ADD therefore requires not only generalization to unfamiliar deepfake methods but also adaptation to newly arriving methods without forgetting earlier ones.

Existing research addresses this challenge through complementary data- and method-centric approaches. Benchmark development, including ASVspoof 2019, the ADD challenge, ASVspoof 5, CodecFake, and AT-ADD, among others, broadens the available coverage of attacks and recording conditions \citep{asv19,add2022,asv5,codecfake,atadd}. On the method side, approaches such as domain generalization aim to learn domain-invariant representations from limited training data, as exemplified by ASDG \citep{asdg} and related domain generalization approaches \citep{ocsoftmax,acs,lsr_lsa}. However, a fixed training set cannot cover continually emerging forgery methods, motivating continual adaptation to new attacks while retaining previously acquired knowledge. Meanwhile, advances in pretrained audio models have substantially improved ADD performance. However, further scaling does not necessarily yield comparable gains in robustness under distribution shift, as recent scaling experiments demonstrate \citep{addscaling}. Generalization-oriented methods cannot guarantee reliable detection of unseen forgery methods, nor do they explicitly address how to incorporate newly emerging attacks without forgetting previously acquired knowledge. This motivates continual adaptation alongside generalization to unfamiliar generators.

Continual learning offers a natural framework for addressing this need: incorporating newly emerging attacks while retaining previously acquired knowledge. It enables learning from successive tasks without repeatedly retraining on the complete historical dataset. General approaches such as EWC and OWM protect earlier knowledge through parameter regularization or constrained updates \citep{ewc,owm}. DFWF pioneered continual learning for ADD by combining knowledge distillation with real-embedding alignment \citep{dfwf}. RAWM subsequently introduced adaptive weight modification and previous-model output regularization, followed by RWM and RegO, which further account for real--fake distribution differences and parameter-region importance \citep{rawm,rwm,rego}. Yet two key questions remain: \emph{how should continual ADD tasks be organized to reflect emerging deepfake generation methods, and how can a detector acquire new deepfake knowledge while preserving previously learned real and fake knowledge?} Unlike conventional class-incremental recognition, the output classes in continual ADD remain real and fake; what evolves is the distribution within those classes. Learning a new task must not sacrifice the ability to detect previously encountered attacks.
\begin{figure}[t]
	\centering
	\includegraphics[width=\linewidth,trim=1.0bp 0.2bp 1.7bp 2.0bp,clip]{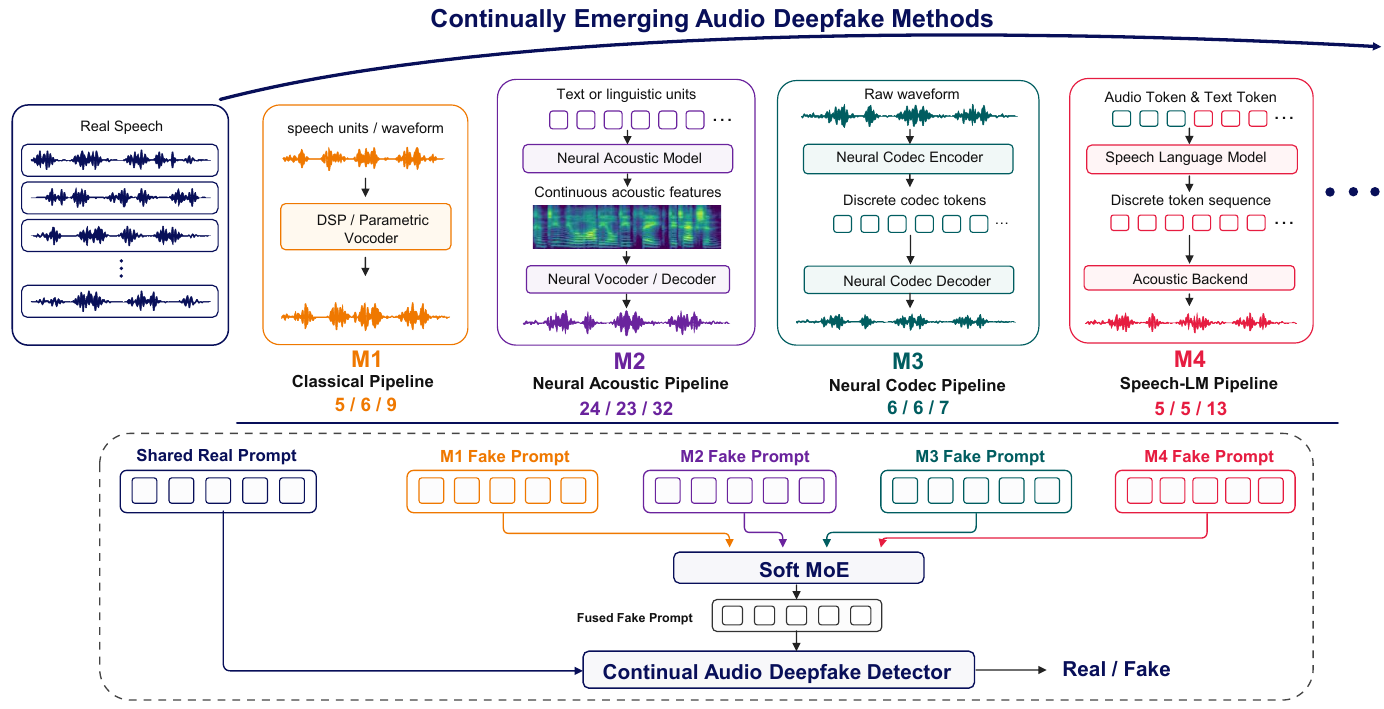}
	\caption{Overview of mechanism-incremental audio deepfake detection. RF-Prompt preserves shared real-speech knowledge while expanding fake experts for emerging deepfake mechanisms. Counts below M1--M4 indicate the numbers of generators in the training, development, and evaluation splits.}
	\label{fig:concept}
	\vspace{-2\baselineskip}
\end{figure}

\textbf{Rethinking task organization.}
Prior continual ADD evaluations, including EVDA as used by RegO, commonly organize learning tasks by dataset \citep{evda,rego}. However, new deepfake generators do not necessarily arrive with new real-speech domains: in practical detector updates, an existing database may already contain abundant real speech from known sources, while additional deepfake methods continue to emerge. The objective is to incorporate these new attacks without losing detection capability on previously observed real and fake speech. Dataset boundaries may obscure this objective because different datasets can share generation mechanisms and one dataset can contain several. We therefore propose the \textbf{real-anchored mechanism-incremental (RAMI) protocol}, which organizes fake samples by generation mechanism against a recurring mixed-domain real-speech background (Figure~\ref{fig:concept}). This design aligns task construction with the need to learn new deepfake mechanisms while reusing available real-speech knowledge, consistent with prior work on shared real characteristics and diverse fake representations \citep{ocsoftmax,asdg,lsr_lsa}. RAMI and four comparison protocols use identical training, development, and evaluation pools to investigate how these task-organization choices affect continual-learning performance.

\textbf{Asymmetric prompt adaptation.}
This updating scenario calls for preserving reusable real-speech knowledge while accommodating increasingly diverse deepfake mechanisms. Updating a shared representation alone risks overwriting earlier knowledge, whereas learning each new task independently limits transfer between related deepfake methods. Prompt-based continual learning offers a parameter-efficient way to allocate shared and task-specific adaptation capacity \citep{l2p,dualprompt,rainbow}. We therefore propose \textbf{RF-Prompt (real--fake prompt learning)}, which protects a shared real prompt and expands task-specific fake experts. Each new expert inherits historical knowledge and learns a complementary residual, allowing adaptation without overwriting stored experts. Cosine anchoring stabilizes the shared real prompt, while residual orthogonality encourages distinct adaptation directions. Input-adaptive soft fusion draws on the accumulated experts while keeping the injected prompt length fixed, so additional stored knowledge does not require a longer prompt sequence.

Our contributions are threefold:
\begin{itemize}
	\item \textbf{RAMI protocol.}
	We propose RAMI to model incrementally emerging deepfake mechanisms against a recurring mixed-domain real-speech background. Together with four other protocols, our evaluation covers five representative task organizations for continual ADD over identical training, development, and evaluation pools.
    \item \textbf{Asymmetric continual prompt learning.} RF-Prompt combines cosine-anchored shared real parameters, inherited fake experts with orthogonal residual regularization, and input-adaptive fixed-length fusion. It retains and expands knowledge without storing historical training audio or adding a feature-distillation forward pass.
    \item \textbf{Multi-axis empirical analysis.} We compare continual-learning and adapted prompt baselines, task organizations, component ablations, limited-data settings, and speech backbones. RF-Prompt obtains the lowest final average and pooled EER among the evaluated continual methods on RAMI: 10.110\% and 10.370\%. The five-protocol comparison examines how real-speech arrival and fake-task organization affect RF-Prompt under a fixed global sample budget.
\end{itemize}

\section{Related Work}
\textbf{ADD datasets and generalization.} ASVspoof 2019, ASVspoof 5, ADD, CodecFake, and AT-ADD broaden coverage of deepfake methods and recording conditions \citep{asv19,asv5,add2022,codecfake,atadd}. ASDG and W2V-ASDG aggregate real representations across domains while separating fake representations \citep{asdg,w2vasdg}; one-class and prototype-based approaches similarly exploit compact real structure and diverse fake patterns \citep{ocsoftmax,acs,lsr_lsa}. These findings motivate asymmetric modeling, while RAMI examines how source-domain arrival and mechanism organization affect continual adaptation.

\textbf{Continual ADD.} DFWF combines knowledge distillation with real-embedding alignment \citep{dfwf}. RAWM, RWM, and RegO develop adaptive update constraints and real--fake-aware knowledge protection \citep{rawm,rwm,rego}. Oiso et al.\ use prompt tuning for target-domain adaptation \citep{oisoprompt}. We examine successive adaptation and retention under controlled task organizations, alongside these established approaches.

\textbf{Continual learning and prompt adaptation.} Classical methods such as EWC and OWM protect historical knowledge through regularization and constrained updates \citep{ewc,owm}. Prompt-based methods instead adapt lightweight tokens: L2P retrieves prompts, DualPrompt combines shared and task-specific prompts, and CODA-Prompt composes input-weighted components \citep{l2p,dualprompt,codaprompt}. RainbowPrompt, KA-Prompt, SinglePrompt, and SMoPE further investigate diversity, alignment, sharing, and sparse expert selection \citep{rainbow,kaprompt,singleprompt,smope}. RF-Prompt assigns shared and expanding capacity to real and fake knowledge, respectively. Appendix~\ref{app:related} provides the detailed review.

\section{RAMI: a real-anchored mechanism-incremental protocol}

To reflect detector updates involving new deepfake mechanisms and available real speech from known domains, RAMI combines mechanism-incremental fake tasks with a recurring mixed-domain real background. Four comparison protocols over identical sample pools assess how task organization affects continual learning.

\subsection{Learning setting and locked sample pools}

We consider a sequence of datasets $\mathcal D_1,\ldots,\mathcal D_T$, where each sample contains an utterance $x$ and a binary label $y\in\{\mathrm{real},\mathrm{fake}\}$. When learning Task $t$, training accesses the current training partition but no historical training audio. After learning each task, we evaluate the detector on all task evaluation partitions. Task boundaries are available during training; the task identity is unavailable at inference.

The benchmark draws from ASVspoof 2019 LA, ASVspoof 5 Track 1, CodecFake, and the clean AT-ADD Track 2 Speech subset. We first lock training, development, and evaluation membership, then change only task assignment. The training and development pools each contain 9,600 real and 9,600 fake utterances; the evaluation pool contains 20,000 of each class. All five protocols therefore share 19,200 training, 19,200 development, and 40,000 evaluation samples. Dataset-qualified identities are retained to avoid merging unrelated generator identifiers with the same spelling.

\subsection{Generation mechanisms}\label{sec:mechanisms}

We organize fake speech into four groups according to the generation pathway. \textbf{M1, Classical Pipeline}, covers classical signal processing, traditional parametric vocoding, and speech-unit concatenation. \textbf{M2, Neural Acoustic Pipeline}, covers neural acoustic modeling or vocoding that produces continuous acoustic features, latent representations, or waveforms. \textbf{M3, Neural Codec Pipeline}, encodes and reconstructs existing speech through a neural codec. \textbf{M4, Speech-LM Pipeline}, generates a target speech sequence using a speech language model, followed by acoustic realization.

M3 reconstructs existing speech, whereas M4 generates a new speech-token sequence. A downstream neural vocoder does not change either assignment to M2. Source composition and generator assignments appear in Appendix~\ref{app:data}.

\begin{figure}[t]
\centering
\includegraphics[width=\linewidth,trim=2.0bp 0.0bp 2.3bp 0.0bp,clip]{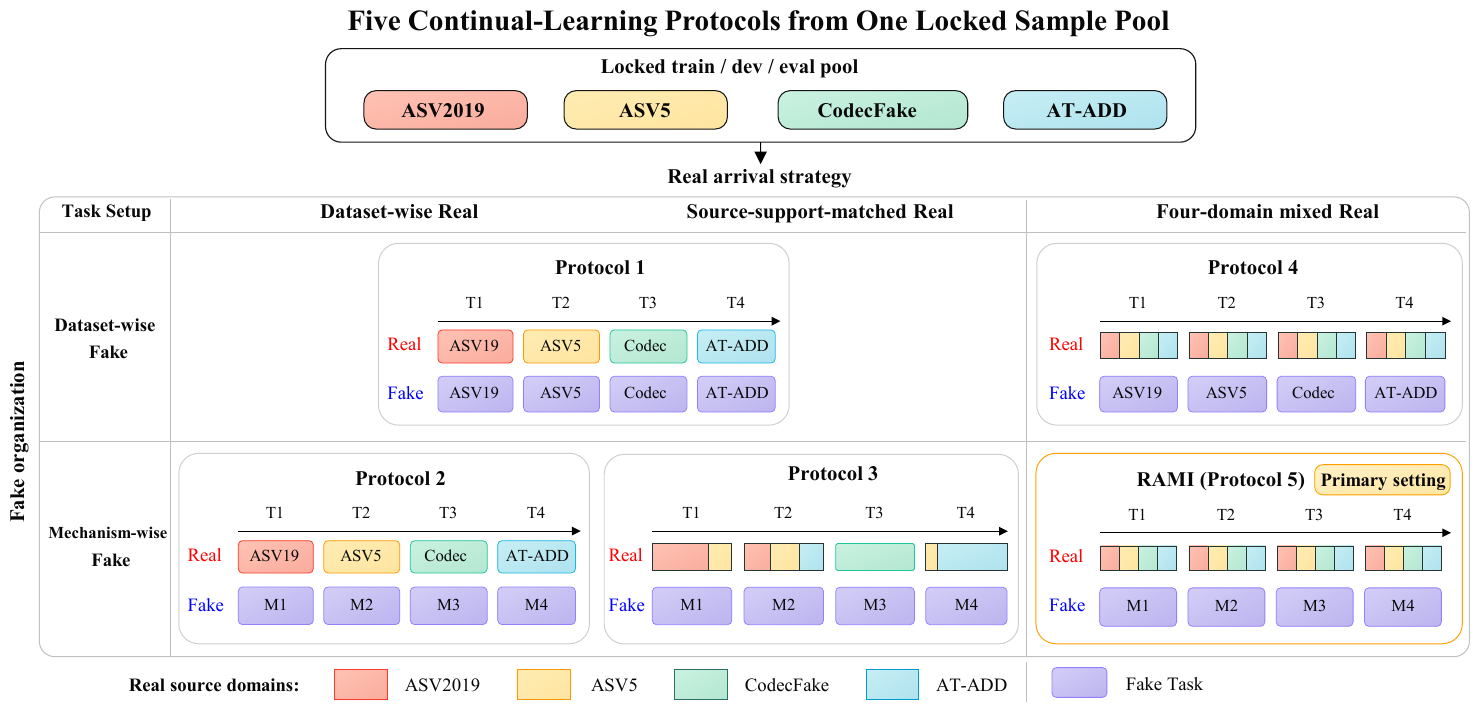}
\caption{Five task organizations from identical train, development, and evaluation pools. Columns vary real arrival, while rows vary fake organization. RAMI (Protocol 5) is the proposed protocol; recurring real domains use disjoint utterances across tasks.}
\vspace{-1\baselineskip}
\label{fig:protocols}
\end{figure}

\subsection{Five task organizations}

Figure~\ref{fig:protocols} crosses real arrival with fake organization. Protocol 1 groups both classes by dataset; Protocol 2 retains dataset-wise real arrival and groups fake speech by mechanism. Protocol 3 matches real-source support to each mechanism task. Protocols 4 and 5 supply a recurring four-domain real mixture, with dataset-wise and mechanism-wise fake tasks, respectively. RAMI is Protocol 5. Its training tasks each contain 600 fresh real utterances from each domain and 2,400 fake utterances from one mechanism. These real utterances are disjoint across tasks; full split counts are in Appendix~\ref{app:data}.

\section{RF-Prompt: Asymmetric Continual Prompt Learning}

To preserve shared real-speech knowledge while learning new deepfake patterns, RF-Prompt protects a shared real prompt and expands fake experts through historical knowledge inheritance and complementary residual learning. Input-adaptive fusion combines these experts into a fixed-length prompt sequence.

\begin{figure}[t]
\centering
\includegraphics[width=\linewidth,trim=0.0bp 1.6bp 0.5bp 2.0bp,clip]{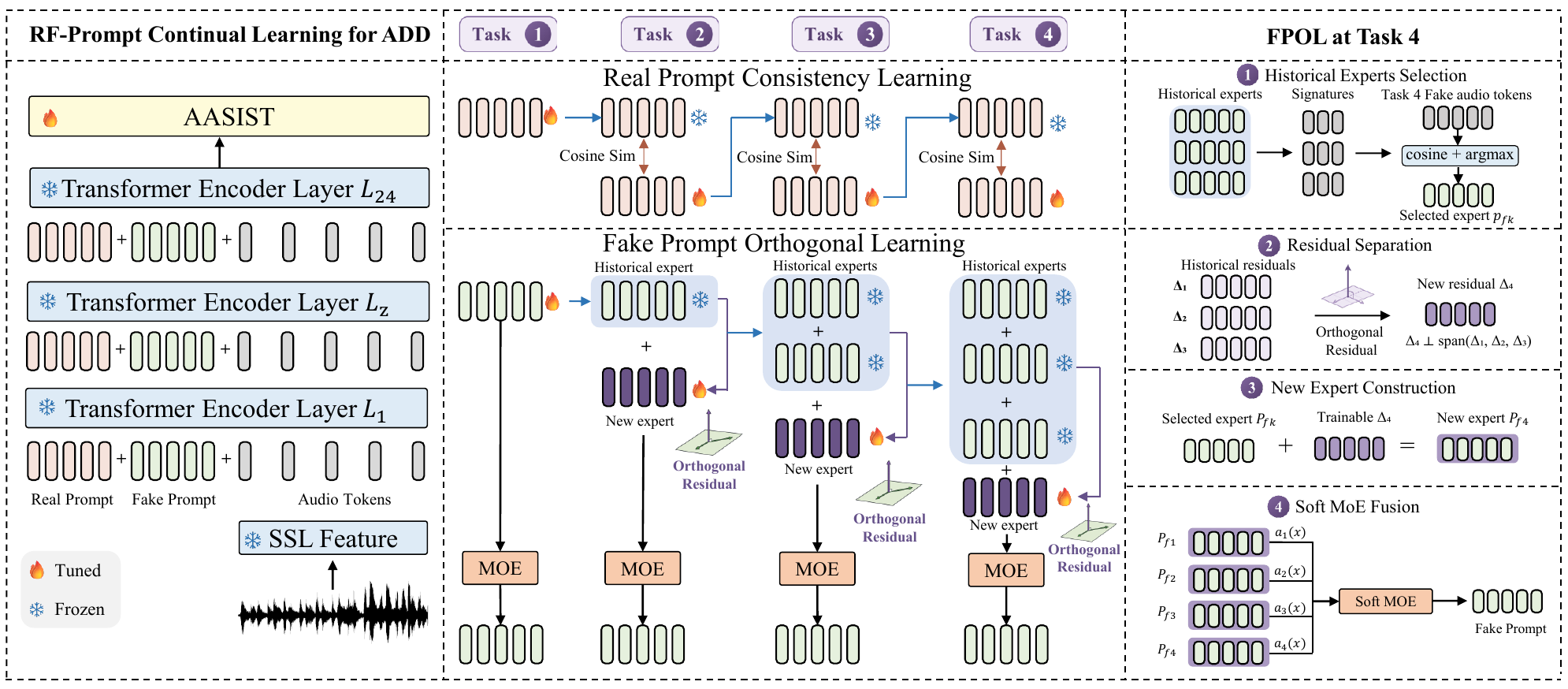}
\caption{Overview of our proposed RF-Prompt. Left: the overall audio deepfake detection pipeline. Middle: continual learning of the shared real prompt and task-specific fake experts. Right: fake expert selection, residual learning, expert construction, and soft fusion at Task 4 as an example.}
\label{fig:model}
\vspace{-1\baselineskip}
\end{figure}

\subsection{RF-Prompt Framework}

Figure~\ref{fig:model} shows the framework. The frozen speech backbone has $L$ transformer layers of width $d$. At task $t$, layer $l$ maintains a shared real prompt $\mathbf R_t^l\in\mathbb R^{m\times d}$ and complete fake experts $\{\mathbf P_{f,k}^l\}_{k=1}^{t}$ of the same shape. Defaults are $L=24$, $d=1024$, and $m=5$. For audio tokens $\mathbf H_t^{l-1}(x)$, the prompted transformation is

\begin{equation}
\mathbf H_t^l(x)=\operatorname{Audio}\!\left(
\mathcal E_l\left([
\mathbf R_t^l;\widetilde{\mathbf P}_{f,t}^l(x);\mathbf H_t^{l-1}(x)
]\right)\right),
\label{eq:1}
\end{equation}

where $\widetilde{\mathbf P}_{f,t}^l(x)$ is the fused fake prompt and $\operatorname{Audio}$ retains only the audio-token outputs. Prompts are introduced anew at each layer rather than accumulating in sequence length. A trainable AASIST backend maps the final audio representation to binary logits.

The shared real prompt and backend remain trainable; historical fake experts are frozen. The real/fake names indicate the intended allocation of shared and expanding capacity, rather than class-exclusive supervision: both classes contribute to cross-entropy, and both prompt types participate in every prediction. The real tokens are not part of the fake mixture.

\subsection{Consistency learning for the shared real prompt}

Sharing real parameters allows knowledge reuse but also exposes earlier real-speech knowledge to later updates. At the beginning of Task $t>1$, we retain the previous selected real prompt $\bar{\mathbf R}_{t-1}$ as a detached reference. We anchor corresponding token directions across the first $K$ layers:

\begin{equation}
\mathcal L_{\mathrm{real}}=
\frac{1}{Km}\sum_{l=1}^{K}\sum_{j=1}^{m}
\left[1-\cos\left(\mathbf R_{t,j}^l,
\operatorname{sg}(\bar{\mathbf R}_{t-1,j}^l)\right)\right].
\label{eq:2}
\end{equation}

Here $\operatorname{sg}$ denotes stop-gradient. We use $K=L$ by default and set the loss to zero for Task 1. This parameter-level constraint protects shared knowledge without historical audio or a teacher-feature forward pass.

\subsection{Orthogonal learning for fake prompts}

\textbf{Selecting transferable knowledge.} Let $\mathbf q(x)\in\mathbb R^d$ be the normalized temporal mean of frozen SSL front-end representations, and let $\mathbf s_k$ be the normalized mean of expert $k$'s prompt tokens across layers. At task initialization, we average queries from current-task fake speech to obtain $\bar{\mathbf q}_t$ and select
\[
k^\star=\arg\max_{k<t}\cos(\bar{\mathbf q}_t,\mathbf s_k),
\qquad \mathbf B_t=\operatorname{sg}(\mathbf P_{f,k^\star}).
\]
This task-level inheritance occurs once; input-adaptive fusion below instead routes each utterance over all available experts. The new expert is initialized with a small perturbation projected away from the historical complete-Prompt space; Appendix~\ref{app:initialization} provides the construction.

\textbf{Separating residual knowledge.} During training, the complete current expert is optimized while its base stays fixed. Its residual is

\begin{equation}
\boldsymbol\Delta_t^l=\mathbf P_{f,t}^l-\mathbf B_t^l.
\label{eq:7}
\end{equation}

Let $\mathbf Q_{\Delta,t}^l$ and $\mathbf Q_{\Delta,<t}^l$ be row-orthonormal factors of the current residual and concatenated historical residuals, obtained by reduced QR of their transposes. If their row counts are $r_t$ and $r_h$, we penalize their normalized overlap:

\begin{equation}
\mathcal L_{\mathrm{orth}}=
\frac{1}{L}\sum_{l=1}^{L}
\frac{\|\mathbf Q_{\Delta,t}^l(\mathbf Q_{\Delta,<t}^l)^\top\|_F^2}
{r_t r_h}.
\label{eq:8}
\end{equation}

Historical factors are detached, and the loss is zero for Task 1. Regularizing residuals preserves the inherited base while encouraging complementary new directions. Orthogonality is a soft training objective.

\subsection{Input-adaptive fusion and optimization}

For any input utterance, we compute weights over all available complete fake experts:

\begin{equation}
\alpha_k(x)=
\frac{\exp(\cos(\mathbf q(x),\mathbf s_k)/\tau)}
{\sum_{r=1}^{t}\exp(\cos(\mathbf q(x),\mathbf s_r)/\tau)},
\qquad
\widetilde{\mathbf P}_{f,t}^l(x)=\sum_{k=1}^{t}\alpha_k(x)\mathbf P_{f,k}^l,
\label{eq:9}
\end{equation}

We use $\tau=0.1$ and share routing weights across layers. Fusion combines complete experts and requires no inference-time task identity.

The task objective combines binary cross-entropy with the two retention terms:

\begin{equation}
\mathcal L_t=
\mathbb E_{(x,y)\sim\mathcal D_t}\mathcal L_{\mathrm{cls}}(x,y)
+\lambda_{\mathrm{real}}\mathcal L_{\mathrm{real}}
+\lambda_{\mathrm{orth}}\mathcal L_{\mathrm{orth}},
\label{eq:10}
\end{equation}

We use $\lambda_{\mathrm{real}}=1$ and $\lambda_{\mathrm{orth}}=0.1$. Five fused fake tokens and five shared real tokens keep the injected length fixed at ten per layer. The expert bank grows with tasks. Appendix~\ref{app:implementation} gives the checkpoint-selection and training procedure.

\section{Experiments}

We evaluate RF-Prompt against continual-learning and adapted prompt baselines, and examine the effect of task organization. Ablations, limited-data experiments, and alternative speech backbones further assess the contributions and applicability of the method.

\subsection{Experimental setup}

We use the frozen XLS-R 300M model\footnote{\url{https://huggingface.co/facebook/wav2vec2-xls-r-300m}} with a trainable AASIST backend, 50 epochs per task, batch size 32, and Adam with a cosine learning-rate schedule. Each task's own development set selects its checkpoint by EER. Table~\ref{tab:1} compares matched continual-learning and adapted prompt baselines; offline joint co-training is a separate reference. Experiments use seed 2026; complete settings appear in Appendix~\ref{app:implementation}.

Let $e_{t,j}$ be the EER on task $j$ after learning task $t$. We report final average EER, $T^{-1}\sum_j e_{T,j}$, and pooled EER obtained from all evaluation scores with a single threshold sweep. Average forgetting is

\begin{equation}
\mathrm{AF}=\frac{1}{T-1}\sum_{j=1}^{T-1}
\left(e_{T,j}-\min_{t\in\{j,\ldots,T\}}e_{t,j}\right).
\label{eq:11}
\end{equation}

For cross-protocol comparisons, pooled EER uses the same 40,000 utterances. We additionally regroup scores into the same four RAMI (Protocol 5) evaluation groups to compute a common average. Native averages from different task partitions are not interchangeable. All results below are single-seed observations; no statistical significance is implied.

\begin{figure}[t]
\centering
\includegraphics[width=.6\linewidth]{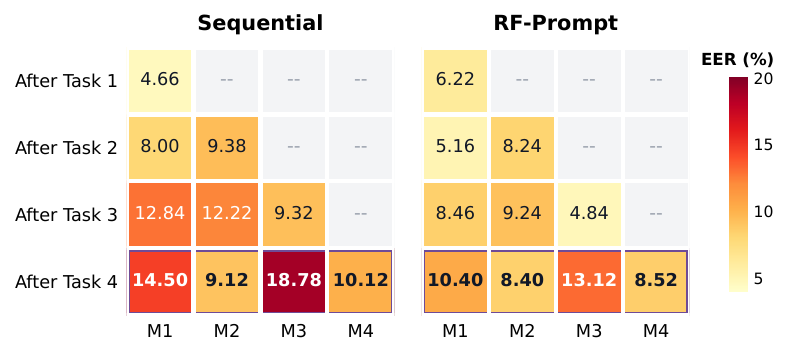}
\caption{Acquired-task EER trajectories of Sequential and RF-Prompt on RAMI (Protocol 5). Rows correspond to the checkpoint after each task and columns to acquired deepfake mechanisms; lower EER is better. Purple boxes mark the final checkpoint. Complete trajectories for all continual methods appear in Appendix~\ref{app:trajectories}.}
\label{fig:retention}
\vspace{-1\baselineskip}
\end{figure}

\subsection{Main comparison and continual retention}

\begin{table}[t]
\centering
\caption{Final results on RAMI (Protocol 5) with XLS-R 300M. EER and AF are reported in \%; lower is better. Category and venue identify each method's original scope and publication. All scores are from our matched experiments; $\dagger$ marks methods adapted to continual ADD. Offline co-training is excluded when identifying the best continual method.}
\label{tab:1}
\scriptsize
\setlength{\tabcolsep}{2.7pt}
\resizebox{\textwidth}{!}{%
\begin{tabular}{lllrrrrrrr}
\toprule
Method & Category & Venue & M1 & M2 & M3 & M4 & Avg EER & Pool EER & AF \\
\midrule
Sequential & Traditional CL & -- & 14.50 & 9.12 & 18.78 & 10.12 & 13.130 & 13.445 & 6.433 \\
EWC~\citep{ewc} & General CL & PNAS'17 & 13.40 & 8.30 & 17.88 & 9.44 & 12.255 & 12.805 & 5.827 \\
OWM~\citep{owm} & General CL & Nat. Mach. Intell.'19 & 13.80 & 8.68 & 17.64 & 9.90 & 12.505 & 13.080 & 5.993 \\
RAWM~\citep{rawm} & ADD-specific CL & ICML'23 & 10.22 & 8.02 & 14.96 & 11.98 & 11.295 & 11.730 & 5.753 \\
RWM~\citep{rwm} & ADD-specific CL & AAAI'24 & 13.40 & 8.32 & 17.46 & 9.88 & 12.265 & 12.845 & 5.700 \\
RegO~\citep{rego} & ADD-specific CL & AAAI'25 & 14.46 & 8.22 & 17.72 & 9.64 & 12.510 & 13.000 & 6.060 \\
Oiso Prompt$^\dagger$~\citep{oisoprompt} & ADD domain adaptation & Interspeech'24 & 7.34 & 13.46 & 25.92 & 30.10 & 19.205 & 20.350 & \textbf{1.067} \\
SinglePrompt$^\dagger$~\citep{singleprompt} & Prompt-based CL & CVPR Findings'26 & 12.82 & 7.52 & 17.44 & 8.88 & 11.665 & 12.105 & 5.593 \\
KA-Prompt$^\dagger$~\citep{kaprompt} & Prompt-based CL & ICML'25 & 12.00 & 7.80 & 18.66 & 9.30 & 11.940 & 12.415 & 5.440 \\
SMoPE$^\dagger$~\citep{smope} & Prompt-based CL & ICLR'26 & 13.38 & 8.34 & 20.40 & 10.80 & 13.230 & 13.615 & 6.533 \\
RF-Prompt (ours) & ADD prompt-based CL & Ours & 10.40 & 8.40 & 13.12 & 8.52 & \textbf{10.110} & \textbf{10.370} & 4.560 \\
Joint (offline) & Offline reference & -- & 7.20 & 7.56 & 11.52 & 10.60 & 9.220 & 9.420 & --- \\
\bottomrule
\end{tabular}%
}
\vspace{-1\baselineskip}
\end{table}

Our method achieves the lowest average and pooled EER among the continual methods, improving over the strongest prompt-based competitor, SinglePrompt, by 1.555\% and 1.735\%, respectively. It also improves average and pooled EER over EWC by 2.145\% and 2.435\% and reduces AF from 5.827\% to 4.560\%. Compared with the strongest ADD-specific competitor RAWM, average EER decreases from 11.295\% to 10.110\%, pooled EER from 11.730\% to 10.370\%, and AF from 5.753\% to 4.560\%. Oiso Prompt has lower AF but fails to acquire later mechanisms, yielding 25.92\% and 30.10\% EER on M3 and M4; its low forgetting therefore does not indicate stronger overall continual learning. The improvements are not uniform across tasks: several baselines achieve lower M2 EER. Joint co-training remains better in aggregate, with a pooled gap of 0.950\%.

Figure~\ref{fig:retention} compares retention with Sequential. RF-Prompt finishes with 10.40\% EER on M1 and 13.12\% on M3, versus Sequential's 14.50\% and 18.78\%. Its M2 EER recovers from 9.24\% after M3 to 8.40\% after M4, close to 8.24\% immediately after acquisition. Average forgetting decreases from Sequential's 6.433\% to 4.560\%. Complete trajectories for all baselines appear in Appendix~\ref{app:trajectories}.

\subsection{Effect of task organization}

\vspace{-1\baselineskip}
\begin{table}[htbp]
\centering
\caption{Controlled protocol comparison for our RF-Prompt model. The common average uses identical RAMI (Protocol 5) evaluation groups.}
\label{tab:2}
\small
\begin{tabular}{lrrrr}
\toprule
Protocol & real arrival & fake organization & Common Avg EER & Pool EER \\
\midrule
1 & Dataset-wise & Dataset-wise & 16.400 & 16.430 \\
2 & Dataset-wise & Mechanism-wise & 15.675 & 15.900\\
3 & Source-support-matched & Mechanism-wise & 13.230 & 13.110 \\
4 & Four-domain mixture & Dataset-wise & 11.035 & 11.330 \\
5 (RAMI) & Four-domain mixture & Mechanism-wise & \textbf{10.110} & \textbf{10.370} \\
\bottomrule
\end{tabular}
\vspace{-0.5\baselineskip}
\end{table}

The five protocols cover two continual-learning regimes. Protocols 1--3 continually introduce both real and fake distributions, whereas Protocols 4 and 5 expose all four real source domains from the beginning and focus on the continual acquisition of emerging deepfake mechanisms. Accordingly, Protocols 1--3 evaluate joint real--fake continual learning, while Protocols 4 and 5 evaluate deepfake-incremental learning against a known mixed-domain real background.

For RF-Prompt, changing from dataset-wise fake tasks in Protocol 1 to mechanism-wise tasks in Protocol 2, with the same real-arrival strategy, lowers common-average EER from 16.400\% to 15.675\% and pooled EER from 16.430\% to 15.900\%. With mechanism-wise fake tasks retained, source-support-matched real arrival in Protocol 3 lowers these metrics further to 13.230\% and 13.110\%. Under the recurring mixed-domain real background, RAMI attains 10.110\% and 10.370\%, compared with 11.035\% and 11.330\% for Protocol 4. Thus, RAMI gives RF-Prompt its lowest aggregate errors among the five tested organizations. These comparisons characterize task-organization effects under a fixed global sample pool; per-task sizes and class proportions also change with the grouping. RAMI represents the practical setting of available real domains and arriving deepfake mechanisms.

\subsection{Component ablations}

\begin{table}[t]
\centering
\caption{Ablations on RAMI. Each control retains the other selected settings; the initialization control removes only historical-space projection.}
\label{tab:3}
\small
\begin{tabular}{lrrr}
\toprule
Configuration & Avg EER & Pool EER & AF \\
\midrule
RF-Prompt (full) & \textbf{10.110} & \textbf{10.370} & \textbf{4.560} \\
w/o real cosine-anchoring loss & 10.755 & 11.365 & 5.293 \\
w/o residual-orthogonality loss & 12.255 & 12.805 & 7.320 \\
w/o adaptive fusion (uniform mean) & 11.525 & 12.480 & 5.480 \\
w/o orthogonal initialization projection & 11.085 & 11.430 & 5.867 \\
\bottomrule
\end{tabular}
\vspace{-1\baselineskip}
\end{table}

In the uniform-mean control, we replace the input-adaptive weights in Eq.~\ref{eq:9} with $\alpha_k(x)=1/t$ for all available complete fake experts. In the initialization control, we retain historical-expert selection and the residual perturbation scale, but skip removing the perturbation's projection onto the historical complete-Prompt space.

Removing any component increases both average and pooled EER. The largest pooled degradation, 2.435\%, occurs when the residual orthogonal loss is removed while orthogonal initialization is retained. Initialization alone is therefore insufficient in this comparison. Replacing adaptive fusion with a uniform mean increases pooled EER by 2.110\%, supporting input-dependent weighting. Removing real anchoring and initialization projection increases pooled EER by 0.995\% and 1.060\%, respectively. These are controlled single-seed ablations rather than estimates of independent additive component effects.

Appendix~\ref{app:ablations} compares alternative real prompt protection and fake expert constructions.

\begin{figure}[t]
\centering
\includegraphics[width=\linewidth]{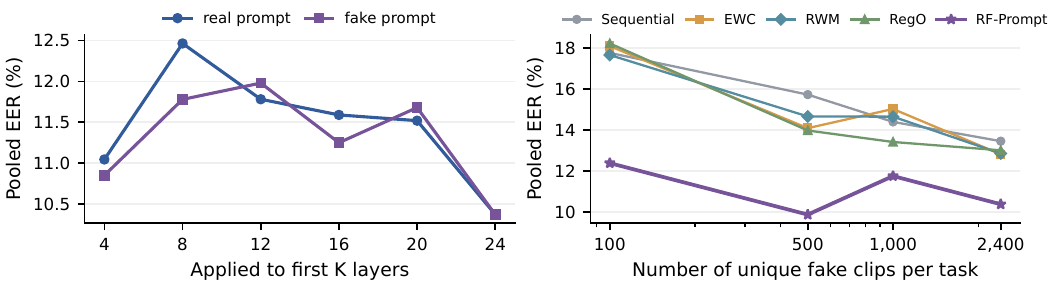}
\caption{Protection-depth sweeps (left) and limited-fake-data adaptation (right). The other constraint remains active across 24 layers; the right axis counts unique fake clips per task.}
\label{fig:sensitivity}
\vspace{-1\baselineskip}
\end{figure}
\vspace{-1\baselineskip}
\subsection{Sensitivity and transfer}
\textbf{Protection depth.} Figure~\ref{fig:sensitivity} (left) varies the first $K$ constrained layers under the selected loss weights, keeping the other constraint active across all 24 layers. Both sweeps are non-monotonic and attain their lowest pooled EER of 10.370\% at $K=24$, consistent with the all-layer configuration.

\textbf{Limited new fake training data.} Figure~\ref{fig:sensitivity} (right) restricts fake training speech to 100, 500, or 1,000 unique utterances per task, retaining 2,400 real utterances. Resampling preserves class balance and optimization steps; development and evaluation sets remain fixed, including 2,400 fake development utterances per task. RF-Prompt outperforms the four compared baselines at every budget. With 100 fake training utterances, pooled EER is 12.375\%, versus 17.645\% for the strongest compared baseline. Appendix~\ref{app:sensitivity} provides details.

\textbf{Transfer across backbones.} RF-Prompt improves average EER, pooled EER, and AF over Sequential on WavLM-Large, W2V-BERT 2.0, XLS-R 1B, and XLS-R 2B, attaining pooled EERs of 12.920\%, 10.040\%, 9.570\%, and 9.550\%, respectively. Appendix~\ref{app:extensions} reports the complete comparisons.

\section{Conclusion}

We introduced RAMI, a real-anchored mechanism-incremental protocol, together with four controlled alternatives over the same sample pools. RF-Prompt achieves its lowest pooled and common-average EER under RAMI among these five organizations. Cosine-protected shared prompts, inherited residual experts, and fixed-length adaptive fusion improve performance under RAMI and in matched low-resource and cross-backbone comparisons. Changing real domains, long task sequences, and unseen generators remain open challenges.

\clearpage
\section*{AI Use Statement}

Generative AI tools were used to assist with manuscript drafting and language polishing. All AI-assisted text was reviewed by the authors, and the experimental results, analyses, and scientific claims were verified by the authors. The authors take full responsibility for the final content of this work.

\bibliography{references}
\bibliographystyle{iclr2027_conference}
\clearpage
\appendix
\FloatBarrier
\section{Benchmark Construction and Data Composition}\label{app:data}
This appendix documents how one fixed sample pool is reorganized into the five continual-learning protocols. We distinguish three levels throughout the construction: the \emph{source dataset} identifies the corpus from which an utterance is drawn, the \emph{generation mechanism} assigns a fake utterance to M1--M4, and the \emph{continual task} determines when that utterance becomes available for learning.

\subsection{Locked sample pools}
The benchmark draws speech from ASVspoof 2019 LA, ASVspoof 5 Track 1, CodecFake, and AT-ADD Track 2 Speech. We use the speech-detection setting and exclude the music, singing, and environmental-sound subsets of AT-ADD. Sample membership is fixed independently for training, development, and evaluation before any continual task is constructed. All five protocols therefore contain the same utterances, labels, audio paths, and split membership; they differ only in how these samples are assigned to Tasks 1--4.

Each split is globally class-balanced. Training and development each contain 9,600 real and 9,600 fake utterances, while evaluation contains 20,000 real and 20,000 fake utterances. Within the real pool, each of the four source domains contributes 2,400 training utterances, 2,400 development utterances, and 5,000 evaluation utterances. The training fake pool contains 2,500 ASVspoof 2019, 800 ASVspoof 5, 2,400 CodecFake, and 3,900 AT-ADD utterances. This locked construction enables task organizations to be compared without changing the global data available to the learner.

\subsection{Five protocol organizations}
The five protocols combine three real-arrival strategies with two fake-task organizations. Dataset-wise arrival introduces samples according to their source dataset; mechanism-wise arrival groups fake samples by M1--M4; mixed-domain real arrival repeatedly draws from all four known real-source domains.

\textbf{Protocol 1} organizes both real and fake speech by source dataset. \textbf{Protocol 2} retains dataset-wise real arrival but organizes fake speech by generation mechanism. \textbf{Protocol 3} uses mechanism-wise fake tasks and matches the real-source composition of each task to the source support of its fake data: within each split and source domain, its fixed real budget is allocated across tasks in proportion to that domain's fake counts, with rounding remainders assigned by largest fractional share. \textbf{Protocol 4} combines an equal four-domain real mixture with dataset-wise fake tasks. \textbf{RAMI (Protocol 5)} combines the same recurring real mixture with mechanism-wise fake tasks. Dataset-wise tasks follow ASVspoof 2019, ASVspoof 5, CodecFake, and AT-ADD, whereas mechanism-wise tasks follow M1--M4.

Table~\ref{tab:appcounts} reports the resulting number of utterances in every task and split. Protocols 1 and 4 use 2,400 real utterances per training/development task and 5,000 per evaluation task; their total task sizes vary with the dataset-wise fake allocation. Protocols 2 and 5 contain 2,400 real and 2,400 fake utterances per training/development task, and 5,000 of each class per evaluation task. Protocol 3 fixes the same per-task fake budget as Protocols 2 and 5, while its real count varies according to source-support matching.

\begin{table}[!htbp]\centering\scriptsize\setlength{\tabcolsep}{4pt}
\caption{Total utterances per task. Each cell lists Train/Dev/Eval counts.}\label{tab:appcounts}
\begin{tabular}{lrrrr}
\toprule
Protocol & Task 1 & Task 2 & Task 3 & Task 4 \\
\midrule
1 & 4,900/4,505/9,832 & 3,200/3,535/8,525 & 4,800/4,800/10,000 & 6,300/6,360/11,643 \\
2 & 4,800/4,800/10,000 & 4,800/4,800/10,000 & 4,800/4,800/10,000 & 4,800/4,800/10,000 \\
3 & 4,704/5,526/10,600 & 5,819/5,019/10,380 & 4,800/4,800/10,000 & 3,877/3,855/9,020 \\
4 & 4,900/4,505/9,832 & 3,200/3,535/8,525 & 4,800/4,800/10,000 & 6,300/6,360/11,643 \\
5 (RAMI) & 4,800/4,800/10,000 & 4,800/4,800/10,000 & 4,800/4,800/10,000 & 4,800/4,800/10,000 \\
\bottomrule
\end{tabular}
\end{table}

Protocols 4 and 5 represent continual adaptation with an existing multi-domain real-speech pool and newly arriving deepfake data. Each task receives a disjoint real subset containing 600 utterances from each source domain in training and development, and 1,250 from each domain in evaluation. RAMI uses this recurring real background to support continual acquisition of mechanism-organized fake knowledge.

\FloatBarrier
\subsection{Generation mechanisms and RAMI composition}
We assign each fake generator to M1--M4 according to its dominant generation pathway, following Section~\ref{sec:mechanisms}. Speech-LM generation of a new speech-token sequence belongs to M4, whereas neural-codec reconstruction of existing speech belongs to M3; a downstream decoder or vocoder does not change these assignments. For hybrid classical/neural acoustic systems at the M1/M2 boundary, the waveform-generating or waveform-enhancing stage determines the assignment. For example, ASVspoof 2019 A07 uses WORLD synthesis followed by a WaveCycleGAN2 neural post-filter and is assigned to M2~\citep{asv19database}. Table~\ref{tab:catalog} reports the complete assignments.

Table~\ref{tab:fakecounts} gives the source composition of the four mechanism-defined fake tasks under RAMI. Every mechanism contributes 2,400 fake utterances to training and development and 5,000 to evaluation, while its source-dataset composition follows the available samples in the locked pool.

\begin{table}[!htbp]\centering\footnotesize\setlength{\tabcolsep}{3.5pt}
\caption{Source-dataset composition of the four mechanism-defined fake tasks under RAMI.}\label{tab:fakecounts}
\begin{tabular}{l p{.22\linewidth} p{.22\linewidth} p{.30\linewidth}}
\toprule
Mechanism & Train & Dev & Eval \\
\midrule
M1 & ASV2019: 2,400 & ASV2019: 2,000; ASV5: 400 & ASV2019: 3,890; ASV5: 1,110 \\
M2 & ASV2019: 100; ASV5: 800; AT-ADD: 1,500 & ASV2019: 105; ASV5: 735; AT-ADD: 1,560 & ASV2019: 942; ASV5: 2,030; AT-ADD: 2,028 \\
M3 & CodecFake: 2,400 & CodecFake: 2,400 & CodecFake: 5,000 \\
M4 & AT-ADD: 2,400 & AT-ADD: 2,400 & ASV5: 385; AT-ADD: 4,615 \\
\bottomrule
\end{tabular}\end{table}

The numbers of represented generators in training/development/evaluation are 5/6/9 for M1, 24/23/32 for M2, 6/6/7 for M3, and 5/5/13 for M4. Generator identifiers are indexed jointly by source dataset and code, so Table~\ref{tab:catalog} lists the complete assignment under both fields. In the locked training pool, M1, M3, and M4 each draw from one source dataset, while M2 spans three; mechanism and source are therefore not fully crossed. M1 development and evaluation also include ASVspoof 5, and M4 evaluation includes ASVspoof 5, testing cross-source transfer within those mechanism groups. M1 development EER thus includes this source shift during checkpoint selection.

\begin{table}[!htbp]
\centering\scriptsize\setlength{\tabcolsep}{3pt}\renewcommand{\arraystretch}{0.92}
\caption{Generation-mechanism assignment by source dataset.}\label{tab:catalog}
\begin{tabular}{l l p{.69\linewidth}}
\toprule

Source & Mechanism & Generator IDs in assignment catalog \\

\midrule

ASV2019 & M1 & A02, A03, A04, A05, A06, A11, A13, A14, A16, A17, A18, A19 \\

ASV2019 & M2 & A01, A07, A08, A09, A10, A12, A15 \\

ASV5 & M1 & A12, A19, A20 \\

ASV5 & M2 & A01, A02, A03, A04, A05, A06, A07, A08, A09, A10, A11, A13, A14, A15, A16, A17, A18, A21, A22, A23, A24, A25, A26, A27, A28, A30, A31, A32 \\

ASV5 & M4 & A29 \\

AT-ADD & M2 & BigVGAN, DiffGANTTS, DiffSpeech\_fastdiff, E2TTS, F5TTS, FastDiff, FastPitch\_fastdiff, FastSpeech2\_fastdiff, GlowTTS, GradTTS, HiFiGAN, Kokoro, MBMelGAN, MelGAN, MeloTTS, OpenVoice2, ParallelWaveGAN, PortaSpeech\_normal\_fastdiff, SeedVC1, StarGANv2VC, StyleMelGAN, StyleSpeech, StyleTTS2, Tacotron2, VITS, WaveGlow, WaveNet, prodiff\_teacher\_fastdiff \\

AT-ADD & M4 & ChatTTS, CosyVoice, CosyVoice2, CosyVoice3, FireRedTTS2, Fish, GPTSoVITS, IndexTTS1, IndexTTS1.5, IndexTTS2, Llasa1B, Llasa3B, Llasa8B, ParlerTTSmini, SparkTTS, StepAudioTTS, Tortoise \\

CodecFake & M3 & F01, F02, F03, F04, F05, F06, F07 \\

\bottomrule

\end{tabular}
\end{table}

\FloatBarrier
\section{Implementation and Optimization Details}\label{app:implementation}
\subsection{Training configuration}
RF-Prompt freezes the SSL backbone and trains the AASIST backend, shared real parameters, and the current fake expert. Previous complete fake experts and inherited bases are detached. Both classes contribute to binary cross-entropy. The real consistency term operates directly on parameters and is sample-independent, despite its real-retention motivation. No historical audio is retained, and no teacher-feature forward pass is used by the selected prompt-cosine configuration.

\begin{table}[htbp]\centering\small\setlength{\tabcolsep}{4pt}
\caption{Selected XLS-R 300M configuration.}\label{tab:hyperparams}
\begin{tabular}{p{.47\linewidth}p{.45\linewidth}}
\toprule

Setting & Value \\

\midrule

Sample rate / input length & 16 kHz / 64,600 samples \\

Tasks / epochs per task & 4 / 50 \\

Batch size / random seed & 32 / 2026 \\

Optimizer & Adam \\

Adam betas / epsilon & (0.9, 0.999) / 1e-8 \\

Weight decay & 5e-4 \\

Initial / minimum learning rate & 1e-4 / 1e-6 \\

Schedule & Cosine decay within each task \\

Training / evaluation workers & 8 / 4 \\

Development evaluation & Every epoch \\

Checkpoint selection & Lowest dev EER; ties: lowest dev loss \\

real / fake tokens per layer & 5 / 5 \\

Prompt dropout & 0.1 \\

cosine coefficient & 1.0 \\

Residual orthogonal coefficient & 0.1 \\

Inherited perturbation scale & 0.1 \\

Routing temperature & 0.1 \\

\bottomrule

\end{tabular}\end{table}

\subsection{Matched baseline settings}
The recorded EWC configuration uses coefficient 100 and importance decay 1.0, without a Fisher-estimation batch cap. OWM uses its projection parameter $\alpha=1.0$. RegO uses importance quantile 0.75 and forgetting threshold 0.1, without an importance-estimation batch cap. The additional LwF reference uses distillation coefficient 1.0 and temperature 2.0. These values are extracted from each run's own configuration; inactive defaults in another method's configuration are not interpreted as active losses. All comparisons use the corresponding frozen-backbone training framework rather than reported scores from the original publications.

SMoPE is adapted from its official prompt-expert implementation to XLS-R attention projections. The audio version uses 25 prefix experts per head, activates the top five in the first six encoder layers, and retains the binary AASIST classifier. Its results are from the same four-task RAMI pool and training budget as the other adapted prompt baselines, not from the original vision benchmarks.

\subsection{Expert construction and tensor dimensions}
For the XLS-R 300M backbone used in our primary experiments, each complete fake expert has shape $24\times5\times1024$, corresponding to 24 transformer layers, five prompt tokens per layer, and a hidden dimension of 1024. Parent selection returns one complete expert with this shape from the historical bank, rather than compressing fifteen historical tokens into five. At Task 4, the historical bank has shape $3\times24\times5\times1024$; its three averaged signatures have shape $3\times1024$. A current-task fake query has dimension 1024 and selects a single inherited expert. The trainable residual has the same shape as this expert. Their elementwise sum is the new complete expert.

At inference, an utterance query produces four softmax weights at Task 4. For a minibatch of size $b$, a layer's bank of shape $4\times5\times1024$ is combined into $b\times5\times1024$ effective fake tokens. The five real tokens are injected separately. Routing uses all complete experts, including those not selected as the current expert's base. Parent selection at task initialization and input-dependent soft routing are therefore different operations.

\subsection{Orthogonal initialization and residual regularization}\label{app:initialization}
\textbf{Initializing new directions.} For each layer, let $\mathbf Q_{P,<t}^l$ contain orthonormal rows obtained from the historical complete Prompt bank. Starting from a random row-normalized matrix $\mathbf U_t^l$, we remove its projection onto the historical space and orthonormalize the remaining rows:

\begin{equation}
\mathbf V_t^l=\mathbf U_t^l-
\mathbf U_t^l(\mathbf Q_{P,<t}^l)^\top\mathbf Q_{P,<t}^l,
\qquad
\widehat{\mathbf U}_t^l=\operatorname{RowQR}(\mathbf V_t^l).
\label{eq:5}
\end{equation}

The initialization scales each new direction relative to its inherited token:

\begin{equation}
\mathbf P_{f,t,j}^{l,\mathrm{init}}=
\mathbf B_{t,j}^l+
\epsilon\,\max(\|\mathbf B_{t,j}^l\|_2,\delta)
\widehat{\mathbf U}_{t,j}^l,
\qquad \epsilon=0.1,
\label{eq:6}
\end{equation}

with a small numerical floor $\delta$. Thus the perturbation is small relative to the inherited prompt, rather than having a fixed absolute magnitude. For task 1, we initialize a standalone prompt and set its base to zero.
Let $\operatorname{RowQR}(A)=\operatorname{qr}(A^\top).Q^\top$ using reduced QR. For each layer, the initialization removes a random matrix's projection onto the historical complete-Prompt space, re-orthonormalizes the resulting rows, and scales them relative to the selected base tokens. Training instead applies the orthogonal loss to $\Delta_t=P_{f,t}-B_t$ and stored historical residuals. For Task 1, $B_1=0$, so $\Delta_1=P_{f,1}$. The base remains fixed even if a different expert later receives the largest mixture weight.

If $Q_hQ_h^\top=I$, then $(U-UQ_h^\top Q_h)Q_h^\top=0$ in exact arithmetic. Subsequent normalization within the projected span preserves this property when sufficient rank remains. Finite precision and QR rank deficiency can weaken the numerical interpretation; the implementation does not supply a rank-adaptive guarantee for arbitrarily long task sequences. During optimization, orthogonality is only encouraged by a soft loss, not enforced by repeated exact projection. Independent parameter directions also need not produce independent output features.

The task-level parent query uses at most 512 current-task fake training examples, matching the recorded initialization sample cap. For smaller low-resource pools, this cap does not imply 512 unique examples. Expert signatures are normalized averages over layers and token slots. No separate learned keys are introduced in this response-based routing mode.

\subsection{Task-level training procedure}
Algorithm~\ref{alg:rfprompt} summarizes RF-Prompt optimization across the task sequence. Each task adds one fake expert while updating the shared real prompt and detector backend; previously learned fake experts remain frozen, and development EER determines the checkpoint carried to the next task.
\begin{algorithm}[htbp]
\caption{Task-level training procedure of RF-Prompt}
\label{alg:rfprompt}
\begin{algorithmic}[1]
\Require Task sequence $\{\mathcal D_t\}_{t=1}^{T}$; frozen SSL backbone; real prompt $P_r$; fake expert bank $\mathcal B_f$
\Ensure Selected detector and expanded fake expert bank after Task $T$
\For{$t=1,\ldots,T$}
    \If{$t=1$}
        \State Initialize $P_r$ and $P_{f,1}$; set $B_1=0$ and $\Delta_1=P_{f,1}$
    \Else
        \State Load the best-development checkpoint from Task $t-1$
        \State Detach the preceding $P_r$ as its cosine reference and freeze $\mathcal B_f$
        \State Average current-task fake queries and select the most similar historical expert $B_t$
        \State Initialize $\Delta_t$ by projecting a random perturbation away from the historical complete-expert subspace
        \State Construct the current complete expert $P_{f,t}=B_t+\Delta_t$
    \EndIf
    \For{each training minibatch from $\mathcal D_t$}
        \State Compute input-dependent weights over all complete experts in $\mathcal B_f\cup\{P_{f,t}\}$
        \State Fuse them into five fake tokens per layer and inject them with five shared real tokens
        \State Compute the binary classification loss
        \If{$t>1$}
            \State Add real prompt consistency and fake residual-orthogonality losses
        \EndIf
        \State Update $P_r$, $\Delta_t$, and the detector backend
    \EndFor
    \State Select by current-task development EER, breaking ties by development loss
    \State Evaluate it on all evaluation partitions and append the frozen $P_{f,t}$ to $\mathcal B_f$
\EndFor
\end{algorithmic}
\end{algorithm}

\FloatBarrier
\section{Complete Continual-Learning Results}\label{app:trajectories}
\subsection{Retention matrices on RAMI (Protocol 5)}
Table~\ref{tab:retentionmatrix} reports the complete acquired-task EER trajectories for the sequentially evaluated methods in Table~\ref{tab:1}. Unacquired tasks are omitted because they do not measure retention. The results complement the final aggregate metrics by showing how each method changes after every update. RF-Prompt maintains the strongest final average and pooled EER: its final M3 EER is lower than every continual baseline, while its M2 EER remains close to the value obtained immediately after Task 2. Figure~\ref{fig:retention} presents the Sequential and RF-Prompt trajectories as paired heatmaps. Joint co-training is excluded because it is an offline reference rather than a sequential learner.

\begin{table}[htbp]
\centering
\scriptsize
\setlength{\tabcolsep}{3pt}
\caption{Acquired-task EER trajectories on RAMI (Protocol 5). The entry under ``After Task $t$'' lists EERs for $\mathrm{M1}/\cdots/\mathrm{M}t$ in order; the last two columns report final average and pooled EER. All entries are percentages.}
\label{tab:retentionmatrix}
\begin{tabular}{lccccrr}
\toprule
Method & After Task 1 & After Task 2 & After Task 3 & After Task 4 & Final Avg & Final Pool \\
\midrule
Sequential  & 4.66 & 8.00/9.38 & 12.84/12.22/9.32 & 14.50/9.12/18.78/10.12 & 13.130 & 13.445 \\
EWC         & 4.66 & 8.28/9.68 & 13.58/12.34/9.14 & 13.40/8.30/17.88/9.44 & 12.255 & 12.805 \\
OWM         & 4.66 & 9.34/9.80 & 14.08/12.50/8.80 & 13.80/8.68/17.64/9.90 & 12.505 & 13.080 \\
RAWM        & 4.66 & 6.18/5.10 & 8.54/6.62/6.18 & 10.22/8.02/14.96/11.98 & 11.295 & 11.730 \\
RWM         & 4.66 & 9.24/10.08 & 13.80/12.38/9.10 & 13.40/8.32/17.46/9.88 & 12.265 & 12.845 \\
RegO        & 4.66 & 8.82/10.16 & 14.92/12.88/9.34 & 14.46/8.22/17.72/9.64 & 12.510 & 13.000 \\
Oiso Prompt & 6.50 & 7.96/13.16 & 6.66/12.98/24.04 & 7.34/13.46/25.92/30.10 & 19.205 & 20.350 \\
SinglePrompt & 4.38 & 8.48/9.84 & 12.52/11.00/9.10 & 12.82/7.52/17.44/8.88 & 11.665 & 12.105 \\
KA-Prompt   & 4.56 & 8.28/9.76 & 11.56/10.70/9.78 & 12.00/7.80/18.66/9.30 & 11.940 & 12.415 \\
SMoPE       & 4.86 & 8.24/9.58 & 11.94/11.66/9.32 & 13.38/8.34/20.40/10.80 & 13.230 & 13.615 \\
\textbf{RF-Prompt} & 6.22 & 5.16/8.24 & 8.46/9.24/4.84 & 10.40/8.40/13.12/8.52 & \textbf{10.110} & \textbf{10.370} \\
\bottomrule
\end{tabular}
\end{table}

\subsection{RF-Prompt retention matrices on Protocols 1--4}
\begin{figure}[htbp]
\centering
\includegraphics[width=.92\linewidth]{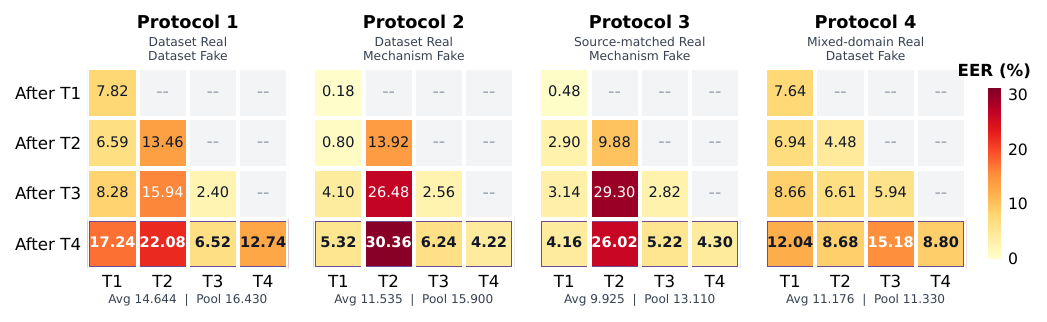}
\caption{RF-Prompt acquired-task EER trajectories on Protocols 1--4. Rows denote the checkpoint after each task, columns denote the task-specific evaluation groups, and purple boxes mark the final checkpoint. All values are EER (\%); lower is better.}
\label{fig:protocolretention}
\end{figure}

Figure~\ref{fig:protocolretention} reports the corresponding RF-Prompt trajectories for the four comparison protocols, following the protocol order and task definitions in Table~\ref{tab:2}. Protocols 1 and 4 use dataset-wise fake tasks, whereas Protocols 2 and 3 use mechanism-wise fake tasks M1--M4. Each panel uses a shared color scale and reports the final native average and pooled EER beneath the matrix.

\raggedbottom
\section{Design exploration for real and fake prompts}\label{app:ablations}
We examine the two asymmetric design choices in RF-Prompt: how shared real knowledge is protected and how new fake experts are constructed. All comparisons in this section use RAMI and the same XLS-R 300M backbone as the primary experiment.

We conduct two complementary design studies. On the real side, we adapt \emph{Shared Prompt Distillation} (SPD) from GAP-Prompt, a prompt-based continual-learning method~\citep{gapprompt}. SPD stabilizes a continually updated shared prompt by maximizing the cosine similarity between intermediate representations produced with the current and previous shared prompts. We include it to examine whether this previously proposed feature-distillation mechanism transfers to continual ADD. In our audio implementation, SPD operates on real samples and aligns hidden representations at the audio-token positions of the protected transformer layers. It therefore provides a feature-level alternative to our parameter-level cosine anchor: SPD requires a reference forward pass with the previous real prompt, whereas Prompt cosine directly compares current and previous real prompt parameters.

\begin{figure}[H]
\centering
\includegraphics[width=\linewidth]{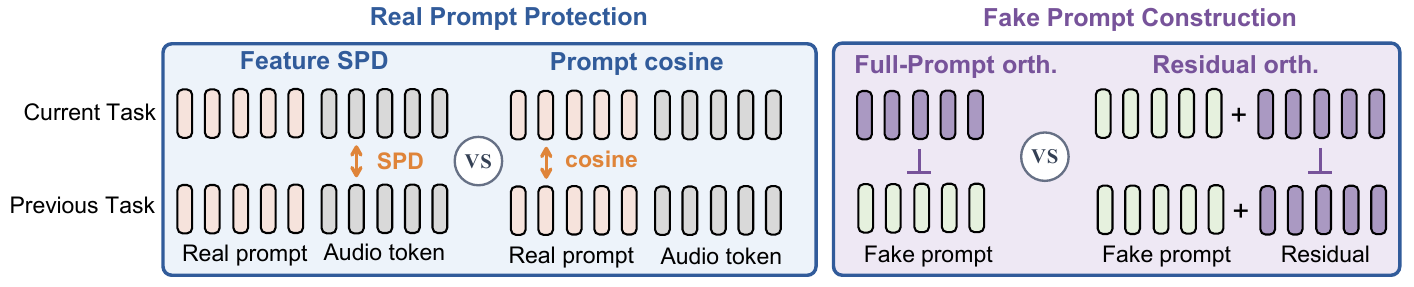}
\caption{Comparison of real prompt protection and fake expert construction. Shared Prompt Distillation (SPD) aligns audio-token features produced with current and previous real prompts; Prompt cosine directly anchors their parameters. Full-Prompt orthogonality separates complete fake experts, whereas residual orthogonality separates only the newly learned residual while retaining the inherited base.}
\label{fig:real_fake_prompt_ablation}
\end{figure}

On the fake side, we use a targeted construction ablation to explain why orthogonality is imposed on residuals rather than complete prompts. The Full-Prompt variant learns a new complete fake prompt directly and regularizes the whole prompt against historical complete experts. RF-Prompt instead decomposes the new expert as $P_{f,t}=B_t+\Delta_t$, where the selected historical expert $B_t$ transfers related deepfake knowledge and only the newly learned residual $\Delta_t$ is separated from historical residuals. Applying orthogonality to the complete expert would also push away the reusable structure intentionally inherited through $B_t$; applying it only to $\Delta_t$ preserves this transferred base while encouraging task-specific knowledge to occupy a complementary direction. Figure~\ref{fig:real_fake_prompt_ablation} illustrates the two real-side alternatives and the fake-side ablation, and Table~\ref{tab:promptdesign} reports their aggregate results. Parameter-level cosine anchoring reduces pooled EER from 11.180\% to 10.370\% relative to SPD, while inherited residual construction reduces it from 12.355\% to 10.370\% relative to Full-Prompt orthogonality.

\begin{table}[H]\centering\small\setlength{\tabcolsep}{4pt}
\caption{Comparison of real prompt protection and fake expert construction on RAMI. All entries are EER or AF (\%).}\label{tab:promptdesign}
\begin{tabular}{llrrr}
\toprule
real protection & fake construction & Avg & Pool & AF \\
\midrule
None & None & 12.600 & 12.020 & 5.393 \\
None & Inherited residual & 10.755 & 11.365 & 5.293 \\
Shared Prompt Distillation (SPD) & Inherited residual & 10.910 & 11.180 & 5.467 \\
Prompt cosine & Full-Prompt orthogonality & 11.795 & 12.355 & 6.140 \\
Prompt cosine & Inherited residual & \textbf{10.110} & \textbf{10.370} & \textbf{4.560} \\
\bottomrule
\end{tabular}
\end{table}

\begin{figure}[H]
\centering
\includegraphics[width=\linewidth]{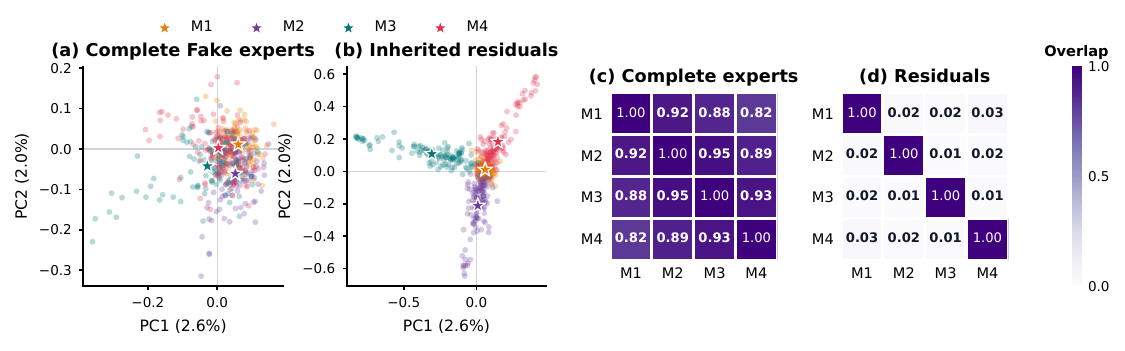}
\caption{Geometry of the learned fake experts on RAMI. Panels (a)--(b) show a joint PCA projection of complete-expert and residual tokens; translucent points are layer-wise prompt tokens and stars are mechanism centroids. Panels (c)--(d) report normalized layer-wise subspace overlap, where lower off-diagonal values indicate stronger separation.}
\label{fig:expertgeometry}
\end{figure}

Figure~\ref{fig:expertgeometry} examines the learned fake-expert geometry at the final RAMI checkpoint. PCA is fitted jointly to row-normalized complete-expert and residual tokens from all transformer layers, with stars denoting task centroids. Because the first two components explain less than 5\% of the variance, the 2D plots are illustrative; the quantitative subspace comparisons support the geometry analysis. The complete experts retain strongly overlapping subspaces because they share inherited knowledge, whereas the residuals occupy distinct directions. We quantify this structure using the normalized layer-wise overlap
\(\frac{1}{m}\lVert Q_i^l(Q_j^l)^\top\rVert_F^2\), where \(Q_i^l\) is an orthonormal basis for expert \(i\) at layer \(l\). Off-diagonal overlap is 0.817--0.951 for complete experts and 0.010--0.030 for residuals. These measurements describe parameter-space overlap under inherited residual construction; they do not directly measure functional or mechanism-specific knowledge separation.

\FloatBarrier
\section{Limited-Data and Backbone Extensions}\label{app:extensions}
\subsection{Unique-fake-utterance budgets}
We restrict each task to 100, 500, or 1,000 unique fake utterances while retaining 2,400 unique real utterances. Sampling with repetition expands the chosen fake subset to 2,400 training rows per epoch. This isolates the amount of unique fake information while preserving class balance, the training-step budget, and the original development and evaluation sets. The 2,400-example setting uses the complete fake task.

\begin{table}[H]\centering\small\setlength{\tabcolsep}{5pt}
\caption{Limited-data results on RAMI. All metrics are percentages; lower is better.}\label{tab:lowfull}
\begin{tabular}{rlrrr}
\toprule
Unique fake/task & Method & Avg EER & Pool EER & AF \\
\midrule
100 & Sequential & 17.325 & 17.750 & 6.887 \\
    & EWC        & 17.705 & 18.080 & 6.707 \\
    & RWM        & 17.075 & 17.645 & 6.067 \\
    & RegO       & 17.765 & 18.210 & 6.040 \\
    & \textbf{RF-Prompt} & \textbf{12.590} & \textbf{12.375} & \textbf{3.907} \\
\midrule
500 & Sequential & 15.220 & 15.715 & 7.367 \\
    & EWC        & 13.720 & 14.080 & 4.693 \\
    & RWM        & 14.220 & 14.650 & 6.460 \\
    & RegO       & 13.680 & 13.965 & 5.380 \\
    & \textbf{RF-Prompt} & \textbf{10.110} & \textbf{9.865} & \textbf{3.907} \\
\midrule
1,000 & Sequential & 13.810 & 14.385 & 6.993 \\
      & EWC        & 14.655 & 15.015 & 7.260 \\
      & RWM        & 14.170 & 14.640 & 7.193 \\
      & RegO       & 12.975 & 13.405 & 5.887 \\
      & \textbf{RF-Prompt} & \textbf{11.390} & \textbf{11.745} & \textbf{5.653} \\
\midrule
2,400 & Sequential & 13.130 & 13.445 & 6.433 \\
      & EWC        & 12.255 & 12.805 & 5.827 \\
      & RWM        & 12.265 & 12.845 & 5.700 \\
      & RegO       & 12.510 & 13.000 & 6.060 \\
      & \textbf{RF-Prompt} & \textbf{10.110} & \textbf{10.370} & \textbf{4.560} \\
\bottomrule
\end{tabular}\end{table}

\subsection{Prompt-token count}
\begin{figure}[H]
\centering
\includegraphics[width=.66\linewidth]{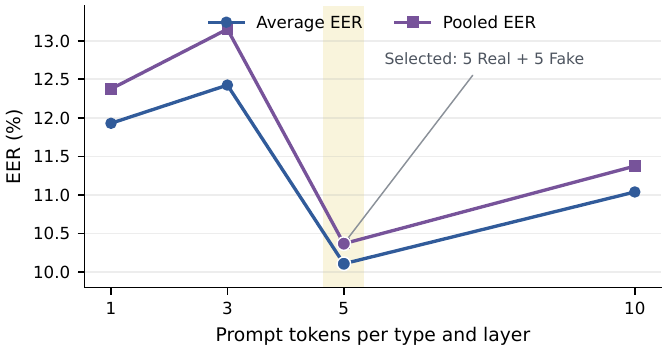}
\caption{Sensitivity to the number of real and fused fake prompt tokens per transformer layer on RAMI. The selected configuration uses five tokens for each prompt type, for ten injected tokens in total.}
\label{fig:promptcount}
\end{figure}

We vary the numbers of real and fused fake prompt tokens injected at each transformer layer while retaining the remaining RAMI training configuration. Figure~\ref{fig:promptcount} shows that the selected allocation of five tokens per prompt type gives the lowest average and pooled EER among the evaluated lengths. The corresponding AF values for 1, 3, 5, and 10 tokens per type are 6.293\%, 7.340\%, 4.560\%, and 6.073\%, respectively.

\subsection{Cross-backbone configuration and results}
\begin{figure}[H]
\centering
\includegraphics[width=\linewidth]{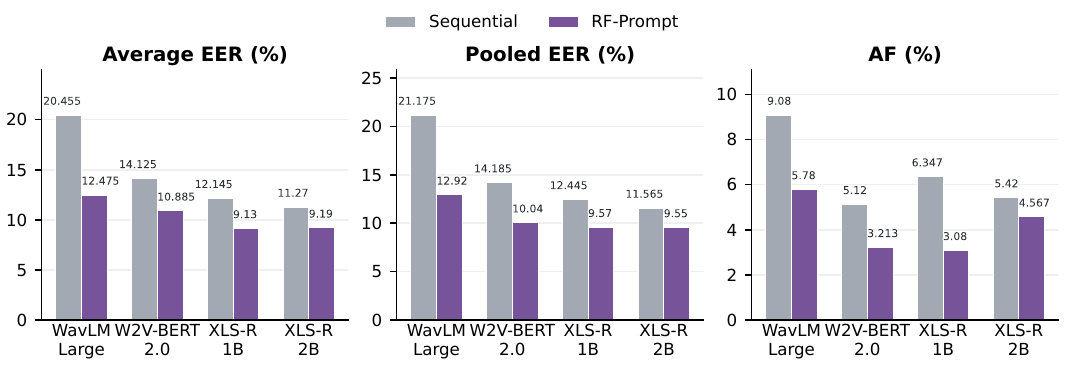}
\caption{Cross-backbone comparison of Sequential and RF-Prompt. Each panel reports one final continual-learning metric after four tasks; lower is better. Values above the bars are percentages.}
\label{fig:sslbackbones}
\end{figure}

The primary XLS-R 300M model and WavLM-Large\footnote{\url{https://huggingface.co/microsoft/wavlm-large}} have 24 layers and hidden width 1024. W2V-BERT 2.0\footnote{\url{https://huggingface.co/facebook/w2v-bert-2.0}} has approximately 600M parameters, while XLS-R 1B\footnote{\url{https://huggingface.co/facebook/wav2vec2-xls-r-1b}} and XLS-R 2B\footnote{\url{https://huggingface.co/facebook/wav2vec2-xls-r-2b}} have 48 layers and widths 1280 and 1920, respectively. Prompt width and backend projection are adapted to each backbone. Every comparison uses four completed tasks, and Sequential and RF-Prompt are evaluated with the same backbone implementation. Figure~\ref{fig:sslbackbones} shows that RF-Prompt consistently improves average EER, pooled EER, and AF across all four alternative backbones.

\FloatBarrier

\FloatBarrier
\section{Extended Related Work}\label{app:related}

We review audio deepfake datasets and detection methods, followed by continual-learning approaches with a focus on prompt-based adaptation.

\subsection{Audio Deepfake Datasets}

Audio deepfake datasets provide complementary coverage of generation methods and recording conditions. ASVspoof 2019 includes logical-access attacks produced by speech synthesis and voice conversion, while ASVspoof 5 extends evaluation to more recent attacks and diverse recording conditions \citep{asv19,asv5}. The ADD challenge complements this benchmark family with detection tasks addressing challenging acoustic conditions and partially manipulated audio \citep{add2022}. CodecFake focuses on audio produced through neural codec reconstruction, broadening coverage beyond conventional synthesis and conversion pipelines \citep{codecfake}. AT-ADD further expands evaluation across speech, singing, music, and environmental sound \citep{atadd}. These resources differ in both real-audio sources and deepfake generation mechanisms. EVDA evaluates continual ADD across eight dataset-defined tasks with changing sources and conditions, and RegO uses that benchmark \citep{evda,rego}. RAMI instead compares real-arrival and fake-organization choices over the same locked sample pool.

\subsection{Audio Deepfake Detection Methods}

\textbf{Domain generalization.}
ASDG aggregates real-speech representations across domains while separating fake representations, and W2V-ASDG integrates this objective with self-supervised speech features \citep{asdg,w2vasdg}. Related approaches encourage compact real representations through one-class learning and adaptive centroid shift \citep{ocsoftmax,acs}, or model fake diversity using multiple prototypes and latent-space augmentation \citep{lsr_lsa}. ADD systems also leverage pretrained XLS-R representations with layer-selective classification \citep{zhang2024audio} and attentive merging of WavLM's multi-layer representations \citep{pan2024attentive}, although scaling alone does not guarantee robustness under distribution shift \citep{addscaling}.

\textbf{Continual learning and domain adaptation.}
DFWF was the first study to apply continual learning to fake-audio detection, combining learning without forgetting with positive alignment of real embeddings \citep{dfwf}. RAWM then adapted weight-modification directions using the real/fake sample ratio and regularized the current detector with outputs from its preceding version \citep{rawm}. RWM and RegO further address continual ADD through real--fake-aware and region-dependent optimization, respectively \citep{rwm,rego}. Oiso et al.\ propose efficient prompt-based adaptation with limited target data \citep{oisoprompt}, focusing on target-domain performance rather than retention across successive tasks. RF-Prompt instead combines shared real-knowledge protection with incremental fake-knowledge expansion.

\subsection{Continual Learning and Prompt-Based Adaptation}

Classical continual-learning methods balance new-task adaptation with the preservation of previously acquired knowledge. EWC penalizes changes to parameters estimated to be important for earlier tasks \citep{ewc}. OWM instead projects updates away from previously learned input subspaces to reduce interference \citep{owm}. These approaches provide general mechanisms for mitigating forgetting through parameter regularization or constrained optimization.

Prompt-based continual learning adapts pretrained models through lightweight trainable tokens. L2P learns a prompt pool and retrieves relevant prompts without requiring task identity at inference \citep{l2p}. DualPrompt combines task-invariant and task-specific prompts to accommodate shared and specialized knowledge \citep{dualprompt}. CODA-Prompt assembles decomposed prompt components with input-conditioned attention weights \citep{codaprompt}. More recent methods investigate how prompt knowledge is reused and integrated: RainbowPrompt evolves task-specific prompts to enhance diversity \citep{rainbow}, while KA-Prompt aligns knowledge components across domain-specific prompts to reduce interference during fusion \citep{kaprompt}. SinglePrompt examines whether prompt selection is necessary in task-free online continual learning \citep{singleprompt}. SMoPE organizes a shared prefix prompt into sparse experts selected for each input \citep{smope}. These studies motivate prompt sharing, expansion, and selection as complementary design choices. RF-Prompt investigates their asymmetric use for real and fake knowledge in continual ADD, where the output classes remain fixed while deepfake generation mechanisms accumulate.

\section{Additional Method Details}\label{app:methoddetails}
\textbf{Selecting transferable knowledge.} Let $\mathbf q(x)\in\mathbb R^d$ be a normalized temporal mean of the frozen SSL front-end representations. For each historical complete expert, define its signature by averaging over layers and token slots:

\begin{equation}
\mathbf s_k=\operatorname{norm}\left(
\frac{1}{Lm}\sum_{l=1}^{L}\sum_{j=1}^{m}\mathbf P_{f,k,j}^l
\right).
\label{eq:3}
\end{equation}

At task initialization, we average up to 512 queries from current-task fake training examples to obtain $\bar{\mathbf q}_t$. The selected historical expert and the frozen inherited base are

\begin{equation}
k^\star=\arg\max_{k<t}\cos(\bar{\mathbf q}_t,\mathbf s_k),
\qquad \mathbf B_t=\operatorname{sg}(\mathbf P_{f,k^\star}).
\label{eq:4}
\end{equation}

This selection occurs once when introducing a task. It differs from per-utterance soft routing at inference and requires no historical audio. We use the complete selected expert as the base, not a concatenation or mean of all historical experts.

\textbf{Initializing new directions.} At each layer, we remove a random perturbation's projection onto the historical complete-Prompt space and re-orthonormalize its rows. The perturbation is scaled by 0.1 times each inherited token's norm before addition to the frozen base. Appendix~\ref{app:initialization} provides the equations and numerical interpretation. Task 1 uses a standalone prompt with zero base.

Our primary model uses frozen XLS-R 300M and a trainable AASIST backend. Audio is loaded at 16 kHz with a fixed input length of 64,600 samples. Each task is trained for 50 epochs with batch size 32, Adam, and a cosine learning-rate schedule from $10^{-4}$ to $10^{-6}$. We evaluate the current task's development set after every epoch, select the lowest development EER, and break ties using development loss. The selected model, rather than the final-epoch model, is evaluated and inherited by the next task. Reported experiments use seed 2026.

We compare Sequential adaptation, EWC, OWM, RWM, and RegO within the same frozen-backbone setup. These are matched implementations in our training framework, not scores copied from their original papers. Offline joint co-training accesses all training tasks simultaneously and is reported separately. It is a useful reference but not a mathematical upper bound.

\section{Protection Depth and Limited-Data Analysis}\label{app:sensitivity}

The left panel of Figure~\ref{fig:sensitivity} examines how many early transformer layers should receive each constraint. We vary the first $K$ layers for real prompt consistency or fake residual orthogonality while keeping the other constraint active across all 24 layers. Under the selected loss weights, both curves achieve their lowest pooled EER of 10.370\% when the corresponding constraint is applied to all 24 layers. We therefore adopt all-layer real prompt consistency and fake residual orthogonality in RF-Prompt.

The right panel evaluates adaptation with limited new fake training speech. Each task retains 2,400 unique real training utterances, while the number of unique fake training utterances is reduced to 100, 500, or 1,000. We resample each selected fake subset to 2,400 training instances per epoch, keeping class balance, optimization steps, and the learning-rate schedule unchanged. The original development pool, including 2,400 fake utterances per task, and evaluation sets are retained. Thus, the varied budget counts unique fake training utterances.

RF-Prompt achieves the lowest pooled EER among all five continual-learning methods at every data budget. Relative to the strongest competing method at each budget, it reduces pooled EER by 5.270\%, 4.100\%, 1.660\%, and 2.435\% with 100, 500, 1,000, and 2,400 unique fake utterances per task, respectively. The particularly large improvement in the 100-shot setting demonstrates that the proposed asymmetric prompts remain effective when adapting to a newly arriving deepfake mechanism with scarce training audio.

\end{document}